# Benchmarking Open-Weight Foundation Models for Global AI Technical Governance

Jason Hung
Internet Society

***Abstract*—**Large language models (LLMs) are increasingly deployed in artificial intelligence (AI) governance analysis across national and international organisations. There is, however, growing evidence that such models produce significantly less accurate responses for countries that are underrepresented in their training data—a pattern described in existing literature as geographic bias. Existing studies examining this phenomenon are subject to three methodological limitations that together undermine their findings: (1) reliance on proprietary systems whose weights are not publicly released, which prevents independent replication; (2) evaluation of model knowledge about years that fall after data collection for model training had concluded, leading to geographic ignorance in addition to the natural limits of each model's knowledge; and (3) use of coarse binary response classification that cannot distinguish models' confident fabrication (HF) from their honest acknowledgement of uncertainty. This study addresses all three limitations by benchmarking four open-weight frontier language models against the Global AI Dataset v2 (GAID v2), a verified ground-truth database of 24,453 indicators across 227 countries published on Harvard Dataverse in January 2026. A total of 18 indicators, mapped to the eight thematic dimensions of the IEEE IRAI 2026 framework, are selected from GAID v2, yielding approximately 2,990 country-metric-year observations across six evaluation years (within the period of 2010–2023). Model responses are classified using a five-category scheme that distinguishes (a) verified accuracy (VA), (b) HF, (c) honest refusal (HR), (d) qualitative hedging (QH), and (e) misattribution (MF). Geographic disparities in accuracy are estimated through mixed-effects logistic regression and difference-in-differences (DiD) analysis. Primary outputs of this study are an IEEE IRAI 2026 conference paper and an openly licensed replication methodology.

*Index Terms*—geographic bias, large language models, AI governance, open-weight models, hallucination, factual accuracy, Global South, IEEE IRAI

## I. INTRODUCTION

### *A. Overview*

The deployment of large language models (LLMs)—artificial intelligence (AI) systems trained on large collections of text to generate fluent, contextually appropriate responses [26]—in policy analysis, public administration, and governance research (e.g., [12]) has accelerated substantially over the past several years. These systems are now routinely used to synthesise cross-national information about AI policy frameworks, research outputs, regulatory developments, and technological infrastructure. The reliability of such applications depends on the factual accuracy of these models, and on whether that accuracy is consistent across the full range of countries and regions about which users may query.

There is, however, a well-documented structural reason to question whether model accuracy is geographically equitably distributed. LLMs are trained on text corpora that overwhelmingly comprise material from wealthy, English-speaking, and geopolitically prominent countries [27]—particularly the United States and Western Europe. Countries in the Global South, and those whose primary languages and governance structures are less represented in international digital archives, feature far less prominently in the data on which these models are trained [28]. As a result, LLMs are likely to be considerably better informed about some countries (especially the Global North) than others (primarily the Global South), and to respond to queries about less-represented and less-resourced countries either (1) with fabricated but confidently stated figures, (2) with generic qualitative descriptions, or (3) with outright refusals to engage with the specific question asked.

This phenomenon—geographic bias in LLM factual knowledge [19]—has significance in the domain of AI governance. Accurate, country-specific data on AI research output, legislative activity, digital infrastructure readiness, workforce equity, and safety-critical compute concentration are foundational to evidence-based AI governance at national and international levels. If the language models that policymakers and analysts increasingly rely upon systematically misrepresent the AI governance landscape of certain regions, this may actively mislead the institutions responsible for governing AI technologies.

### *B. Research Problem*

Despite growing scholarly interest in AI fairness and bias, existing empirical studies of geographic bias in LLMs suffer from three major methodological limitations that restrict the conclusions that can be drawn from them. First, most studies (e.g., [14]) use proprietary models—systems whose trained parameters are not publicly released and whose outputs may change without notice as the provider updates the underlying model—barring independent researchers from replicating reported findings. Second, several studies evaluate model knowledge about years that fall after data collection for model training had concluded, resulting in generating refusals and errors that are a function of temporal knowledge limits rather than geographic bias [18]. Third, most studies (e.g., [16]) classify model responses into a binary correct-or-incorrect scheme, or at best a three-category scheme, failing to distinguish between a model that honestly acknowledges it does not know a specific value and one that fabricates a confident but incorrect statistic—a distinction with substantial implications for how geographic bias should be understood and measured.

### C. Research Aims

This project aims to produce a rigorous, reproducible, and geographically comprehensive assessment of factual accuracy and geographic disparity across four open-weight frontier language models, evaluated against verified AI governance indicators drawn from the Global AI Dataset v2 (GAID v2). It further aims to determine whether geographic disparities in accuracy are moderated by the national origin of each model's developer, and whether such disparities vary systematically across thematic domains of AI governance.

### D. Research Questions

**RQ1.** Do the four open-weight language models examined in this study exhibit statistically significant variation in factual accuracy across geographic regions when queried about AI governance indicators?

**RQ2.** Does the national origin of a model's developer moderate the direction or magnitude of geographic accuracy differentials—specifically, do models developed in the United States and Europe perform differently from those developed in China across the same countries and indicators?

**RQ3.** Do geographic accuracy differentials vary systematically across the eight thematic dimensions of the IEEE International Conference on Responsible Artificial Intelligence 2026 (IEEE IRAI) framework [9]—Ethics, Safety, Security, Transparency, Fairness, Accountability, Regulation, and Adoption?

**RQ4.** To what extent does the proximity of an evaluation year to the end of each model's data collection period independently account for geographic disparities in accuracy, once geographic and thematic factors are controlled for?

## II. LITERATURE REVIEW

### A. Geographic and Cultural Bias in LLMs

There is now a substantial body of empirical literature documenting that LLMs encode the distributional characteristics of their training corpora, producing outputs that systematically favour the cultural, political, and linguistic perspectives of the countries and communities most prominently represented in those corpora [15], [17]. Arora et al. [2] demonstrated that pre-trained language models exhibit weaker alignment with non-Western values compared to Western cultural norms, with systematic differences observable across multiple cultural dimensions. Cao et al. [5] showed that LLMs exhibit what they term geopolitical bias: models trained predominantly on English-language Western web data reproduce North American and European analytical frameworks when addressing questions that could equally be approached from other cultural traditions. Navigli et al. [23] coined the term lingua franca bias to describe the observation that a model's factual accuracy correlates strongly with the linguistic and geopolitical prominence of the subject country in its training data. All these findings provide both theoretical grounding and prior empirical evidence for the geographic bias hypothesis that this project examines directly within the AI governance domain.

### B. The Temporal Limits of Trained Model Knowledge

A fundamental feature of LLMs trained on static text corpora is that they can only report knowledge about events and data that existed at the time their training data was collected. The date at which this data collection concluded—the model's training cutoff—defines an absolute temporal boundary beyond which the model has no direct knowledge. Evaluating model knowledge about years that fall after this boundary does not measure geographic bias; it measures the trivial consequence of asking any system about information it was never given.

A further complication, documented by Cheng et al. [6], is that even within the period nominally covered by a model's training data, the final several months before the cutoff are systematically underrepresented. This is because recently published material has had insufficient time to be widely cited, indexed, and redistributed across the web; the training corpus therefore contains proportionally less material about very recent events than about events of equivalent importance from earlier periods. In consequence, model performance on queries about events near the end of the training period is meaningfully lower than performance on queries about earlier periods, even though the nominal training cutoff suggests such knowledge should be available. This study accounts for this pattern by designating years 2010 to 2019 as the primary evaluation window—where all four models' training knowledge is expected to be most complete—and treating 2022 and 2023 as secondary periods to examine whether proximity to the end of the training period independently affects observed accuracy differentials (i.e., RQ4).

### C. Open-Weight Models and Scientific Reproducibility

The majority of high-profile LLM evaluations in the social-scientific literature have relied on proprietary, closed-source systems (e.g., [1])—models whose trained parameters are held by a commercial provider, accessed exclusively via a programming interface, and potentially updated without documentation or versioning, meaning that a study's findings cannot be independently replicated once the provider has modified the underlying model. Kapoor et al. [11] identify this as a threat to reproducibility in AI evaluation research and advocate for the use of open-weight models—systems whose trained parameters are published and made freely available for download, allowing any researcher to deploy the identical model version used in the original study. Biderman et al. [4] demonstrate that open-weight model evaluations produce more stable and replicable findings than API-based evaluations of equivalent proprietary systems.

This project exclusively employs open-weight models with documented version identifiers, all suitable for local deployment on standard academic computing infrastructure. All four models are available on the Hugging Face Model Hub under publicly documented licences; their parameter weights are fixed and will not change after the study is conducted. This design provides the reproducibility guarantees that proprietary API-based evaluations cannot.

### D. Hallucination and Factual Accuracy in Language Models

Ji et al. [10] provide a comprehensive taxonomy of what is commonly called hallucination in natural language generation—the production of responses that are fluent and apparently confident but factually incorrect. They distinguish intrinsic hallucination (responses that contradict verifiable source material) from extrinsic hallucination (responses that fabricate facts not contained in any source the model was trained on), and identify geographic sparsity in training data as a principal driver of extrinsic hallucination for factual queries about specific countries and regions. The methodological implication is that studies using binary correct-or-incorrect response classification combine fabricated confident responses with honest acknowledgements of uncertainty; distinguishing these two response patterns is essential for understanding the nature of geographic bias. This project addresses this through the five-category taxonomy described in Section III-F.

*E. AI Governance Frameworks and the IEEE IRAI Thematic Dimensions*

The OECD AI Principles [24] represent the most widely adopted international framework for trustworthy AI governance, establishing normative expectations around inclusive growth, human rights, transparency, safety, and accountability. The IEEE IRAI [9] operationalises these principles into eight thematic dimensions—Ethics, Safety, Security, Transparency, Fairness, Accountability, Regulation, and Adoption—which provide the thematic architecture for this project's indicator selection.

The GAID v2, the ground-truth reference database published on Harvard Dataverse in January 2026 and used in this study, draws on 11 internationally validated data sources covering, for example, AI research output, patent activity, governance infrastructure, compute capacity, workforce equity, and digital skills. A notable finding of the data preparation process is that dedicated Ethics measurement instruments are entirely absent from GAID v2 for evaluation years 2010–2022, as the Global Index on Responsible AI (i.e., the primary data source for Ethics data in the dataset) and comparable instruments were not published until 2024. The two Ethics indicators retained in this study (*AI_Benefits* and *AI_Nervous*) are both derived from the Ipsos survey wave of 2023 and cover 34–35 countries. This constitutes a structural limitation of current global AI ethics measurement infrastructure, and is itself a finding of scientific significance that will be discussed in the final paper.

*F. Positioning of This Study Relative to Prior Work*

This project builds directly on—and substantially improves upon—the completed but methodologically flawed Apart Research hackathon study [8], which conducted a preliminary geographic bias audit of a single small language model (Llama-3 8B, queried via 1,704 prompts about 213 countries in the year 2025). Three methodological flaws identified by peer reviewers of that study—(1) querying model knowledge of a year beyond the model's training period, (2) using a coarse three-category response taxonomy, and (3) employing a single model of limited capability with no geopolitical comparator—are addressed comprehensively in the present design. Relative to the broader literature on geographic bias, this study is distinctive in its focus on the AI governance domain, its use of a purpose-built ground-truth dataset verified across 227 countries, its geopolitically balanced model selection, and its adoption of a causal identification strategy—difference-in-differences (DiD) estimation—that allows attribution of observed accuracy differentials to geographic rather than temporal factors.

## III. METHODOLOGY

*A. Methodology Overview*

The evaluation proceeds through six sequential stages. First, 2,990 country-indicator-year observations with confirmed numeric values are extracted from GAID v2, drawn across the 18 retained indicators and six evaluation years (2010, 2013, 2016, 2019, 2022, 2023). Second, three query variants are generated for each observation: a direct numeric query (Variant 1), a comparative query embedding the target country in a regional context (Variant 2), and a contextual query framing the indicator within a governance narrative (Variant 3). In total, this study produces 8,970 unique prompt-observation pairings. Third, each prompt is submitted to all four models via their respective API endpoints at temperature zero to maximise response determinism, yielding 35,880 model responses in total. Fourth, raw text responses are collected and stored with source metadata. Fifth, each response is scored under the five-category rule-based classifier—Verified Accuracy (VF), Confident Fabrication (HF), Honest Refusal (HR), Qualitative Hedging (QH) and Misattribution (MF)—using a fully automated, offline procedure requiring no human annotation. Sixth, the classified response set is passed to three statistical procedures—mixed-effects logistic regression, DiD estimation, and principal component analysis (PCA)—and results are disseminated via this conference paper and an open-source methodology release.

*B. Ground-Truth Data: GAID v2*

The GAID v2 contains 259,546 rows of data covering 24,453 unique indicators, 227 countries, and years from 1998 to 2025. The dataset draws on 11 curated international sources covering, for example, AI research output (Stanford AI Index via Scopus), patent activity (WIPO), AI compute and model development (Epoch AI), national governance infrastructure (World Bank GovTech Maturity Index), AI policy and regulation (Stanford AI Index), digital skills and workforce readiness (Coursera), and public attitudes towards AI (Ipsos / Stanford AI Index). Each row of the dataset records the verified value for a given indicator in a given country in a given year. This study treats these values as ground truth against which model responses are evaluated.

The 18 indicators selected for evaluation (detailed in Section IV-A) were identified through a three-stage screening process: (i) thematic mapping, in which candidate indicators were assessed for relevance to each of the eight IEEE IRAI 2026 dimensions; (ii) coverage analysis, retaining only indicators with verified values for at least ten countries across at least one of the six evaluation years; and (iii) redundancy screening, in which indicators whose values were too closely correlated with another selected indicator were excluded. The redundancy threshold was a Spearman rank-order correlation coefficient exceeding 0.90 across the 2022 cross-section. To verify that the retained 18 indicators measure sufficiently distinct constructs, PCA was conducted on the 2022 cross-section of the eight indicators with coverage of ten or more countries in that year. There, six components are required to explain 90 percent of total variance, confirming that the indicator set captures multiple conceptually independent dimensions.

*C. Model Selection*

Four open-weight frontier language models—systems whose trained parameters are made publicly available for download and local deployment, distinguishing them from proprietary API-only systems—are evaluated. The selection achieves geopolitical balance between two models from Western developers (the United States and the European Union) and two models from Chinese developers, therefore enabling analysis of whether the national origin of a model's developer moderates geographic accuracy differentials.

Several notes on model selection are elaborated. For Meta [21], Llama 4 Maverick is selected over the larger Llama 4 Behemoth (which remains in training as of March 2026 and is not publicly available) and over the earlier Llama 3.3 70B (a smaller, single-model-architecture predecessor). Llama 4 Maverick was accessed via OpenRouter (meta-llama/llama-4-maverick). A small number of early queries were routed through Together AI's FP8 endpoint before the provider was standardised to OpenRouter for the remainder of the evaluation run. For Alibaba Cloud, Qwen3-235B-A22B is selected in preference to the more recently released Qwen3.5 (February 2026, 397 billion parameters) on grounds that Qwen3.5 had been publicly available for fewer than four weeks at the time of this study's design and has not yet accumulated an independently verified benchmark record; Qwen3-235B-A22B, released in April 2025, is a well-characterised model with extensive third-party evaluation data [25]. Specifically, Qwen3-235B-A22B was served via Together AI (model ID: Qwen/Qwen3-235B-A22B-fp8, FP8 quantised deployment). A subset of queries was routed to the Instruct-2507-tput endpoint during an API availability period; both endpoints serve the same underlying model weights at reduced precision. For DeepSeek AI, the March 2025 updated release of DeepSeek-V3 (designated DeepSeek-V3-0324) is used, which incorporates improved post-training refinements relative to the original December 2024 release whilst retaining identical architecture and model weights [7]. DeepSeek-V3-0324 was accessed via Together AI's OpenAI-compatible endpoint (deepseek-ai/DeepSeek-V3). Together AI served the DeepSeek-V3-0324 release (March 2025) at the time of evaluation. DeepSeek-R1 (January 2025), a reasoning-oriented variant that generates extended chain-of-thought deliberation before producing responses, is excluded because its response structure is architecturally incompatible with the structured factual query format used in this study. Moreover, Mistral Large 3 was accessed via the Mistral AI API endpoint mistral-large-latest. At the time of evaluation, this endpoint resolved to the Mistral Large 3 675B Instruct deployment (model card: mistral-large-3-675b-instruct-2512).

Table 1 summarises the four models selected in this study.

**TABLE 1**

*Four open-weight frontier language models evaluated in this study*

| Model | Developer | Architecture | Training Data End Date |
|---|---|---|---|
| Llama 4 Maverick | Meta AI (US) | MoE—17B active / 400B total | August 2024 [20] |
| Mistral Large 3 | Mistral AI, France | MoE—~41B active / 675B total | By late 2025 (estimated) [22] |
| Qwen3-235B-A22B | Alibaba Cloud (China) | MoE—22B active / 235B total | By mid 2025 (estimated) [3] |
| DeepSeek-V3-0324 | DeepSeek (China) | MoE—37B active / 671B total | July 2024 [13] |

*D. Evaluation Years and Temporal Design*

Six evaluation years are used: 2010, 2013, 2016, 2019, 2022, and 2023. These years were selected because: (1) all four models were trained on data that encompasses each of these years, meaning the models cannot be expected to lack knowledge of events from these periods on temporal grounds alone; (2) the period selected for evaluation spans 13 years of AI governance development, enabling longitudinal trend analysis; and (3) they correspond to years in which the key ground-truth indicators in GAID v2 are most extensively populated, particularly WIPO patent publication data which covers all six years.

Years 2010 to 2019 are designated the primary analysis window. These years precede the period in which rapid growth in global AI policy activity began (approximately 2020 onwards) and fall well within the period for which all four models' training data is expected to be most thoroughly represented. Years 2022 and 2023 are retained as a secondary analysis period. These years fall within each model's training period but are closer to the end of that period, where knowledge representation in training data becomes progressively sparser [6]. Their inclusion allows the study to test directly whether proximity to the end of each model's training period independently predicts observed accuracy differentials, after controlling for geographic and thematic factors. Year 2024 is excluded entirely from primary analysis, as it falls after the end of the training period for two of the four models studied (Llama 4 Maverick, DeepSeek-V3-0324).

*E. Query Design*

For each of the approximately 2,990 individual observations in the GAID v2 indicator subset—where each observation is a unique combination of (1) country, (2) indicator, and (3)

year—a structured natural-language query is generated. The base query template takes the form: "*According to [data source], what was [indicator name] for [country] in [year]? Please provide a specific numeric value if the information is available to you.*" Three query variants are produced for each observation, namely a direct numeric request (as above); a comparative framing that asks how the country performed relative to the regional average; and a contextual framing that embeds the query in a brief description of a governance analysis scenario. Such an approach yields a maximum of approximately 35,880 queries across all four models (2,990 observations × 3 variants × 4 models). A minimum viable evaluation run uses one variant per observation, yielding approximately 11,960 queries. All models are queried with a temperature parameter of zero to ensure deterministic and reproducible outputs; a five percent random sample of observations is queried twice to assess within-run response consistency.

*F. Response Classification Scheme*

Each model response is classified into one of five mutually exclusive categories using a rule-based automated classifier. The five categories are summarised in Table 2. Responses containing a numeric figure are compared against the verified GAID v2 value: those within ±10 percent are classified as VF; those outside this threshold are classified as HF. The proportion of responses classified as HF in a given country-indicator combination constitutes the primary outcome variable—the fabrication rate—which serves as the principal empirical measure of geographic bias. A systematically higher fabrication rate for countries in one region than another, after controlling for the indicator and year, is interpreted as evidence of geographic bias in model knowledge. Responses that explicitly acknowledge an absence of specific knowledge are classified as HR; responses providing directional or qualitative descriptions without a numeric commitment are classified as QH; responses about a different country, year, or indicator than queried are classified as MF. A random 5 percent sample of classified responses is manually reviewed by the author to verify classifier accuracy.

**TABLE 2**

*Five-category response classification scheme with definitions and analytical significance*

| **Code** | **Category** | **Definition** | **Significance for Bias Detection** |
|---|---|---|---|
| VF | Verified Accuracy | The model provides a numeric estimate within ±10 percent of the verified GAID v2 value. | Confirms accurate country-level knowledge; provides baseline for geographic accuracy comparisons. |
| HF | Confident Fabrication | The model provides a numeric estimate outside the ±10 percent threshold with apparent confidence; the value is factually incorrect. | Primary indicator of geographic bias: systematic overrepresentation in Global South queries indicates fabricated rather than refused responses. |
| HR | Honest Refusal | The model explicitly acknowledges that it does not know the specific numeric value for the queried country and year. | Indicates calibrated uncertainty; may reflect the relative scarcity of that country's data in training corpora rather than geographic bias per se. |
| QH | Qualitative Hedging | The model provides a directional or comparative description without committing to a specific figure (e.g., 'this country has relatively low AI publication output'). | Secondary bias indicator; qualitative responses cannot be verified against ground truth and may mask underlying factual uncertainty. |
| MF | Misattribution | The model responds with information about a different country, year, or indicator than was queried. | Indicates systematic deflection; may reflect training-data confusion between countries with similar geographic or cultural profiles. |

*G. Statistical Analysis*

*Mixed-effects logistic regression.* The fabrication rate is modelled as a binary outcome (HF versus all other categories) via logistic regression with random intercepts for country, nested within world region. Fixed-effect predictors include world region, model identity, evaluation year, and IEEE IRAI thematic dimension. The model takes the form:

$$logit(P[HF_{ijkmt}]) = \beta_0 + \beta_1 Region_j + \beta_2 Model_k + \beta_3 Year_t + \beta_4 Theme_m + u_j + \varepsilon_{ijkmt}$$

Random intercepts $u_j$ for country capture the baseline accuracy level of each country independently of its regional classification, preventing country-specific factors from confounding the regional fixed-effect estimate.

*DiD estimation.* The geopolitical balance of the model selection—two models from Western developers and two from Chinese developers—enables a DiD design that estimates whether the national origin of a model's developer moderates geographic accuracy differentials. The treatment dimension is developer origin (Western versus Chinese); the comparison dimension is the geographic classification of queried countries (Global North versus Global South). The DiD estimator features the extent to which the Global South accuracy penalty is larger for Western-origin models than for Chinese-origin models, or *vice versa*, above and beyond any general accuracy differences between models.

*PCA on accuracy profiles.* After computing country-level fabrication rates for each of the 18 indicators separately, PCA is applied to the resulting *country × indicator* accuracy matrix. This identifies whether geographic bias is a general, domain-independent phenomenon or whether it manifests primarily within particular IEEE IRAI thematic dimensions—for example, whether Safety and Compute indicators show a different geographic accuracy pattern from Regulation and Adoption indicators.

## IV. DATASET AND EXPERIMENT PLAN

### A. Indicators Selected from GAID v2

The 18 indicators listed in Table 3 were selected from GAID v2 following the three-stage screening process described in Section III-B. Together they yield approximately 2,990 country-metric-year observations (each defined by a unique combination of country, indicator, and year with a verified value in GAID v2) across the six evaluation years. Four candidate indicators were excluded at the redundancy-screening stage: *National AI Cluster Power Capacity* was excluded due to a Spearman correlation of r = 1.00 with *Total Training Compute* in the 2022 cross-section; *World Bank GovTech Digital Citizen Engagement Index* was excluded due to r = 0.955 with the composite *GovTech Maturity Index* (all four GovTech sub-dimension scores are redundant with the composite, r > 0.90); *arXiv AI Publications (Total)* due to r = 0.902 with the broader AI publications count; and *Coursera Data Science Proficiency Score* due to r = 0.822 with the *Coursera Technology Proficiency Score*. Two retained pairs with correlations above 0.70—*Total Model Parameters* with *Total Training Compute* (r = 0.888, n = 12) and *Number of AI Publications* with *Total AI-Related Patent Publications* (r = 0.722 to 0.792 across evaluation years)—are retained on the grounds that they measure conceptually distinct constructs: model scale versus training resource requirements in the first case, and academic research output versus industrial innovation in the second. The Accountability theme retains a single indicator (*AI_Legis*) because the only alternative candidate from GAID v2—*Cumulative AI Legislative Mentions*—has a Spearman correlation of 0.906 with the annual measure, exceeding the redundancy threshold.

**TABLE 3**

*18 indicators selected from GAID v2, mapped to IEEE IRAI 2026 thematic dimensions*

| IRAI Theme | # | Code | Indicator Name | Data Source | Countries | Year Range |
|---|---|---|---|---|---|---|
| Transparency | 1 | *AI_Pubs* | Number of AI Publications: All | Stanford AI Index / Scopus | 171 | 2010, 2013, 2016, 2019 |
| Transparency | 2 | *WIPO_Patents* | Total AI-Related Patent Publications (by applicant country of origin) | WIPO | 115 | All six evaluation years |
| Transparency | 3 | *FWCI* | Field Weighted Citation Impact: All | Stanford AI Index / Scopus | 171 | 2010, 2013, 2016, 2019 |
| Fairness | 4 | *CS_Grad_F* | Share of Computer Science Bachelor's Graduates Who Are Female | Stanford AI Index / OECD | 21 | 2013, 2016, 2019, 2022 |
| Fairness | 5 | *CS_PhD_F* | Share of Computer Science Doctoral Graduates Who Are Female | Stanford AI Index / OECD | 16 | 2013, 2016, 2019, 2022 |
| Adoption | 6 | *WB_GTMI* | World Bank GovTech Overall Maturity Index | World Bank GTMI 2022 | 198 | 2022 |
| Adoption | 7 | *Coursera_Biz* | Coursera Business Skills Proficiency Score | Coursera Global Skills Report | 119 | 2022, 2023 |
| Adoption | 8 | *Coursera_Tech* | Coursera Technology Skills Proficiency Score | Coursera Global Skills Report | 69 | 2022, 2023 |
| Adoption | 9 | *BigData_Biz* | Proportion of Businesses Having Performed Big Data Analysis (all activities, 10+ employees) | OECD.ai | 34 | 2013, 2016, 2019, 2022, 2023 |
| Regulation | 10 | *Nat_AI_Strat* | National AI Strategy Published (binary indicator) | Stanford AI Index | 35 | 2019, 2022 |
| Regulation | 11 | *AI_Bills* | Number of AI-Related Bills Enacted into Law | Stanford AI Index | 124 | 2023 |
| Accountability | 12 | *AI_Legis* | Number of AI References in National Legislative Proceedings (2022 Annual) | Stanford AI Index | 80 | 2023 |
| Safety | 13 | *Train_Compute* | Total Training Compute for AI Models Released (FLOP) | Epoch AI | 22 | All six evaluation years |
| Safety | 14 | *Model_Params* | Total Parameters of AI Models Released | Epoch AI | 25 | All six evaluation years |
| Security | 15 | *ICT_Sec_All* | Proportion of Businesses That Experienced ICT Security Breaches (all activities, 10+ employees) | OECD.ai | 36 | 2013, 2016, 2019, 2022, 2023 |
| Security | 16 | *ICT_Sec_ICT* | Proportion of Businesses in the Information and Communication Sector That Experienced ICT Security Breaches (10+ employees) | OECD.ai | 36 | 2013, 2016, 2019, 2022, 2023 |
| Ethics | 17 | *AI_Benefits* | Proportion of Survey Respondents Agreeing That AI Products Offer More Benefits Than Drawbacks | Stanford AI Index / Ipsos | 34 | 2023 |
| Ethics | 18 | *AI_Nervous* | Proportion of Survey Respondents Agreeing That AI Products Make Them Nervous | Stanford AI Index / Ipsos | 35 | 2023 |

*Note: The total number of country-metric-year observations with verified values is approximately 2,990.*

### B. Known Limitations of the Dataset

There are six limitations of the GAID v2 dataset. First, the ethics indicators are limited to 2023 data. Dedicated AI ethics measurement instruments are absent from GAID v2 for evaluation years 2010–2022. The Global Index on Responsible AI and comparable instruments were first published in 2024 and are therefore excluded from primary analysis. Both Ethics indicators in this study (*AI_Benefits* and *AI_Nervous*) derive from the 2023 Ipsos survey wave and cover only 34–35 countries. This means Ethics-theme findings cannot be compared longitudinally across our full evaluation window. This structural gap is itself a finding of scientific significance: standardised global AI ethics measurement infrastructure did not exist until very recently. The study flags this limitation and analyses the two 2023 Ipsos indicators as a cross-sectional probe of public ethical perception rather than a longitudinal indicator.

Second, only one distinct, non-redundant Accountability indicator could be identified within GAID v2 for the evaluation years. As mentioned, the cumulative legislative proceedings metric—the closest candidate for a second Accountability measure—has a Spearman correlation of 0.906 with the annual measure (*AI_Legis*), exceeding the redundancy threshold of 0.90. Also, as mentioned again, World Bank GovTech sub-dimension scores are similarly redundant (all $r > 0.90$ with the composite index). Structural Accountability metrics (such as the Global Index on Responsible AI) are available only for 2024. Accountability findings are interpreted with appropriate caution given the single indicator; the study discusses this as a data availability constraint rather than a methodological choice.

Third, security indicators are limited to OECD member states. Both Security indicators (Indicators 15 and 16) are derived from OECD.ai and cover approximately 36 predominantly European OECD member states, excluding virtually all of the Global South. Security findings are reported separately as a geographically constrained sub-analysis. Broad cross-national comparisons for this dimension are not drawn.

Fourth, fairness indicators are limited to reporting countries. Gender equity in computer science education (Indicators 4 and 5) covers only 16–21 countries, indicating the limited number of states that report gender-disaggregated graduate data to the OECD. Country scope is acknowledged as a limitation. Findings for this dimension are interpreted within the constraints of available data.

Fifth, the *World Bank GovTech Maturity Index* (Indicator 6) is available only for 2022, limiting longitudinal analysis for the Adoption theme. The 2022 wave is used as a cross-sectional snapshot and combined with other Adoption indicators that cover multiple years.

Lastly, Mistral Large 3 and Qwen3-235B-A22B do not have publicly disclosed, precise end dates of their models' training data. The final months before that training data collection end dates are typically underrepresented in training corpora, as recently published material has had insufficient time to be widely cited and indexed [6]. Evaluation years 2010–2019 are designated the primary analysis window; 2022 and 2023 are treated as a robustness check to test whether proximity to the end of each model's training period independently accounts for observed accuracy differentials.

### C. Volume of Queries

The minimum viable evaluation run—one query variant per observation, across all four models—yields approximately 11,960 queries (2,990 observations × 4 models). The full evaluation run—three query variants per observation—yields approximately 35,880 queries. These figures substantially exceed the initial proposal target of 5,000 queries and are sufficient to support the planned regression-based analysis.

### D. Expected Outputs

*Conference paper*. A full empirical paper submitted to IEEE IRAI 2026, Melbourne, reporting geographic accuracy profiles, model-by-region interaction estimates, DiD findings, and thematic accuracy decompositions across all four models and 18 indicators.

*Open-source methodology.* All query templates, response classification codebooks, scoring scripts, and statistical analysis code released under a Creative Commons Attribution 4.0 licence on GitHub, to enable independent replication and extension to future model releases.

## V. RESULTS

### A. Overall Classification Distribution

Across the full evaluation run of 35,880 queries (2,990 observations × 3 variants × 4 models), 27.4 percent of model responses were classified as VF and 71.8 percent as HF. QH accounted for 0.4 percent of responses and MF for 0.5 percent. HR was effectively absent across all four models (<0.1 percent), indicating that all four models responded with numeric estimates to virtually every structured factual query, irrespective of whether those estimates were within the verified range. The aggregate within-run consistency rate—assessed across a five percent random sample of observations queried twice—was 90.1 percent (Mistral Large 3: 94.9 percent; Llama 4 Maverick: 93.8 percent; Qwen3-235B-A22B: 88.0 percent; DeepSeek-V3-0324: 83.8 percent), confirming that the rule-based classifier produced stable classifications across repeated queries under temperature-zero sampling.

Query framing had a significant and consistent effect on fabrication rates. Direct numeric queries (Variant 1) yielded the highest HF rate (75.9 percent; VF=23.5 percent), comparative framing (Variant 2) reduced the HF rate to 67.0 percent (VF=31.7 percent), and contextual framing (Variant 3) produced an intermediate HF rate of 72.4 percent (VF=27.1 percent). These differences are statistically significant in the mixed-effects regression (Variant 2 versus Variant 1: OR=0.59, $p<0.001$; Variant 3 versus Variant 1: OR=0.82, $p<0.001$), indicating that embedding a query in a regional comparison or governance context reduces the probability of HF, though does not eliminate it.

### B. Model-Level Variation

Fabrication rates varied substantially across the four models, as visualised in Figure 1. Mistral Large 3 recorded the highest VF rates (Global North: 29.7 percent; Global South: 44.5 percent), followed by DeepSeek-V3-0324 (Global North: 25.1 percent; Global South: 37.2 percent), Qwen3-235B-A22B (Global North: 19.6 percent; Global South: 30.6 percent), and Llama 4 Maverick (Global North: 11.0 percent; Global South: 15.9 percent). In terms of overall HF (hallucinatory fabrication) rates, Mistral Large 3 achieved the lowest HF rate (61.2 percent), followed by DeepSeek-V3-0324 (67.3 percent), Qwen3-235B-A22B (73.5 percent), and Llama 4 Maverick (85.1 percent). These differences are statistically significant after controlling for region, year, and thematic dimension: relative to DeepSeek-V3-0324, Llama 4 Maverick shows 2.40 times the odds of fabrication (OR=2.40, $p<0.001$),

Qwen3 1.49 times (OR=1.49, $p<0.001$), and Mistral 0.78 times (OR=0.78, $p<0.001$). The rank ordering is stable across the primary and secondary analysis windows, and model identity is the strongest single predictor of fabrication rate in the regression model.

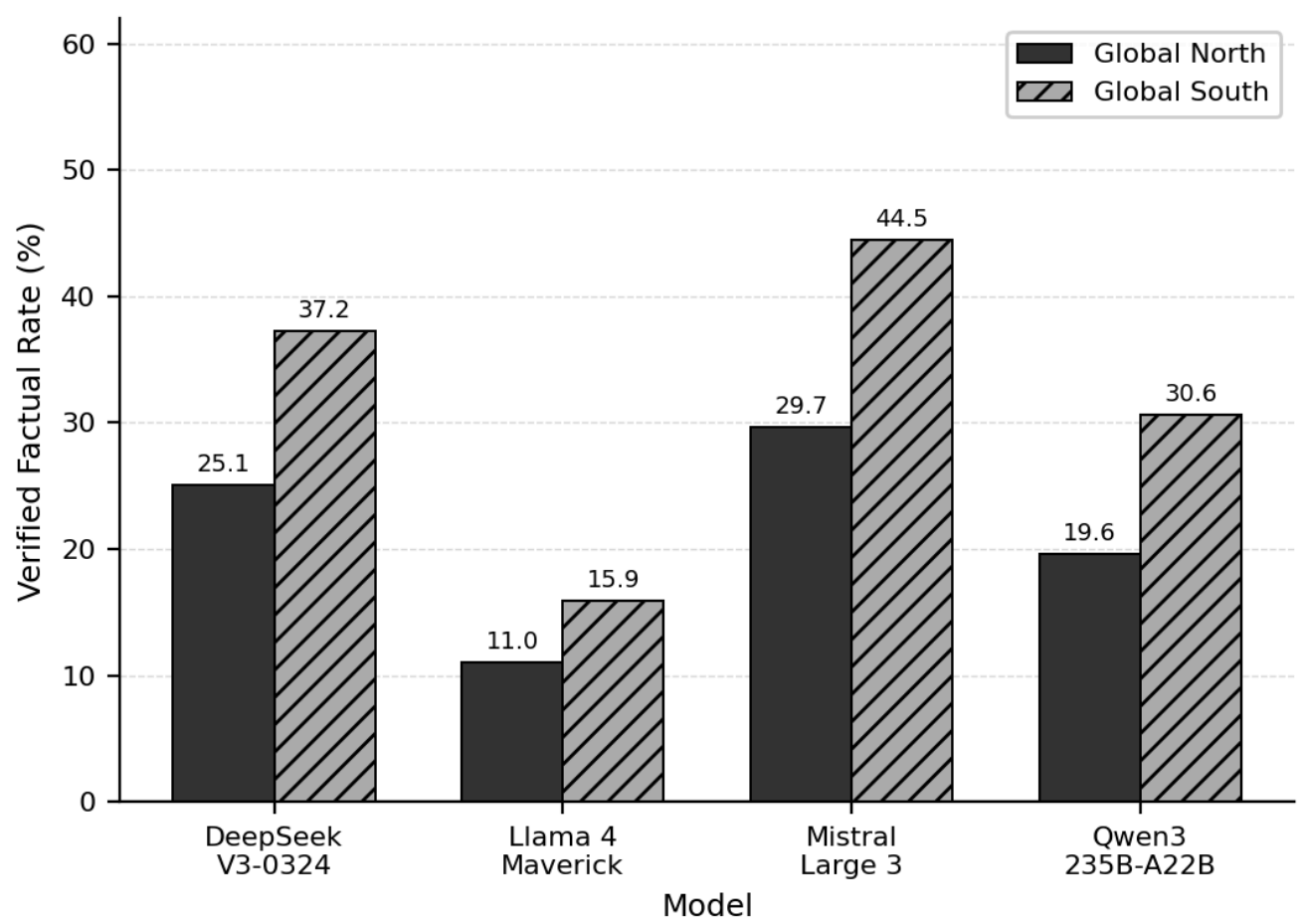

*C. Regional Variation (RQ1)*

Fabrication rates varied across geographic regions in a direction contrary to the primary hypothesis. Africa recorded the highest mean VF rate across all four models (35.1 percent), followed by the Americas (30.2 percent), Asia (28.1 percent), Europe (24.4 percent), and Oceania (23.5 percent). In the mixed-effects logistic regression, Africa showed significantly lower odds of fabrication than Europe, the reference category (OR=0.52, 95% CI [0.38, 0.71], $p<0.001$), as did the Americas (OR=0.66, 95% CI [0.48, 0.91], $p=0.012$). The effect for Asia was in the same direction but did not reach conventional significance at the five-percent level (OR=0.79, 95% CI [0.62, 1.01], $p=0.063$). Oceania did not differ significantly from Europe (OR=0.94, $p=0.850$).

These results indicate that the four models examined in this study do exhibit statistically significant regional variation in fabrication rates (RQ1 confirmed), but the geographic pattern is inverted relative to the dominant hypothesis in the literature. The interpretation of this finding is addressed in Section VI.

*D. Developer Origin and Geographic Disparity (RQ2)*

Figure 2 presents the 2×2 HF rates disaggregated by developer origin (Western versus Chinese) and country income classification (Global North versus Global South). Across both developer groups, Global South countries show noticeably lower HF rates than Global North countries: Chinese-origin models produce HF rates of 65.2 percent for Global South queries versus 77.2 percent for Global North queries (difference: −12.0 percentage points); Western-origin models produce HF rates of 68.8 percent for Global South queries versus 78.8 percent for Global North queries (difference: −10.0 percentage points).

The DiD estimator, which indicates the extent to which the Global South accuracy advantage differs across developer origin groups, yields a coefficient of +2.07 percentage points (OR=1.090, SE=0.049, $z=1.776$, $p=0.076$). This estimate is marginally significant at the ten-percent level. The positive sign indicates that Western-origin models exhibit a slightly smaller Global South accuracy advantage than Chinese-origin models—that is, Chinese-origin models show somewhat greater accuracy consistency between Global South and Global North queries. The developer-origin main effect is negligible and non-significant (OR=1.003, $p=0.941$), confirming that the two groups do not differ in overall fabrication rate but differ modestly in the geographic distribution of fabrication (RQ2: marginal evidence).

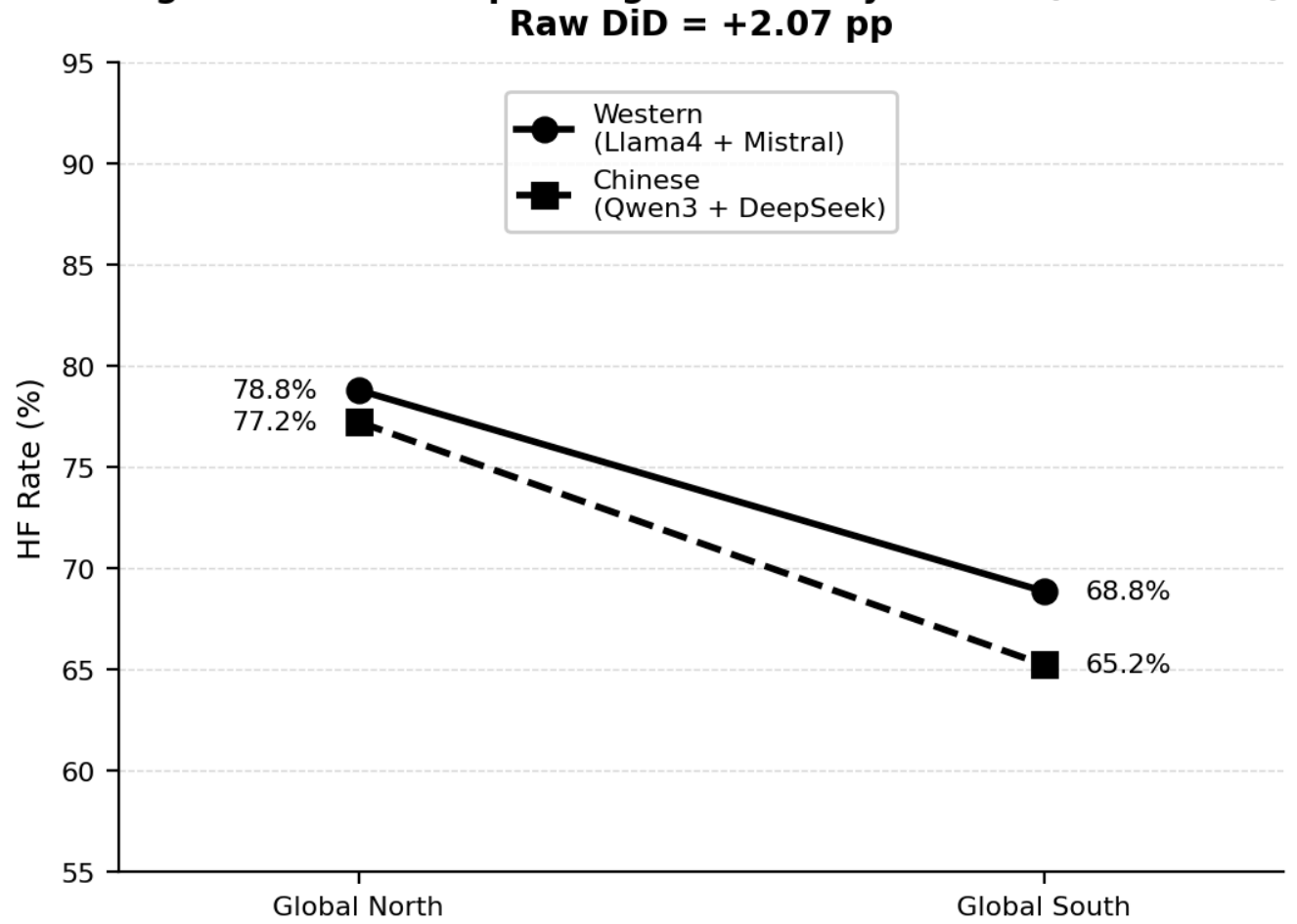

*E. Thematic Variation (RQ3)*

Fabrication rates differed substantially across IEEE IRAI thematic dimensions, as shown in Figure 3. The Safety theme—comprising Total Training Compute (FLOP) and Total Model Parameters Released—recorded a mean VF rate of 1.8 percent across all four models, with Llama 4 Maverick (Safety VF: 0/342=0.00%) and Qwen3-235B-A22B (Safety VF: 1/342=0.29%) achieving (nearly) zero verified responses for Safety indicators. This near-total fabrication shows the extreme magnitude and specificity of AI compute and parameter counts, for which all four models produced plausible-sounding but inaccurate numeric estimates. The Safety effect is the largest thematic effect in the regression (OR=11.93 relative to Transparency, 95% CI [6.16, 23.11], $p<0.001$).

The Regulation theme recorded the highest mean VF rate (42.2 percent), with Mistral Large 3 (54.2 percent) and Qwen3-235B-A22B (53.6 percent) approaching majority accuracy for regulation-related indicators; the regression confirms significantly lower odds of fabrication for Regulation relative to Transparency (OR=0.34, $p<0.001$). Adoption (OR=0.45, $p=0.011$) and Fairness (OR=0.75, $p=0.074$) also showed lower fabrication odds than Transparency, though the Fairness effect is marginal. Security (OR=1.59, $p=0.395$) and Transparency did not differ significantly.

The thematic variation in fabrication rates is not contingent on model identity. All four models rank Safety as their lowest-accuracy theme and Regulation as among their highest, indicating that thematic structure of geographic bias is largely model-invariant.

**Fig. 3. VF Rate (%) by IEEE IRAI Theme and Model (2010–2023)**

| IEEE IRAI Theme | DeepSeek V3-0324 | Llama 4 Maverick | Mistral Large 3 | Qwen3 235B-A22B |
|---|---|---|---|---|
| Regulation | 44 | 17 | 54 | 54 |
| Accountability | 37 | 7 | 46 | 33 |
| Ethics | 32 | 19 | 36 | 36 |
| Transparency | 34 | 15 | 42 | 28 |
| Fairness | 31 | 17 | 38 | 24 |
| Adoption | 29 | 12 | 29 | 15 |
| Security | 14 | 7 | 15 | 19 |
| Safety | 3 | 0 | 4 | 0 |

## F. Principal Component Structure of Geographic Bias (RQ3)

To assess whether geographic bias is a general, domain-independent phenomenon or whether it concentrates within particular thematic dimensions, PCA was applied to the 177-country × 18-indicator matrix of VF rates. The scree plot is shown in Figure 4. Fifteen components are required to explain 90 percent of total variance (PC1: 11.3 percent; cumulative through PC15: 92.9 percent). The high dimensionality of the accuracy matrix indicates that geographic bias is not a single, general factor but is instead distributed across multiple thematic dimensions.

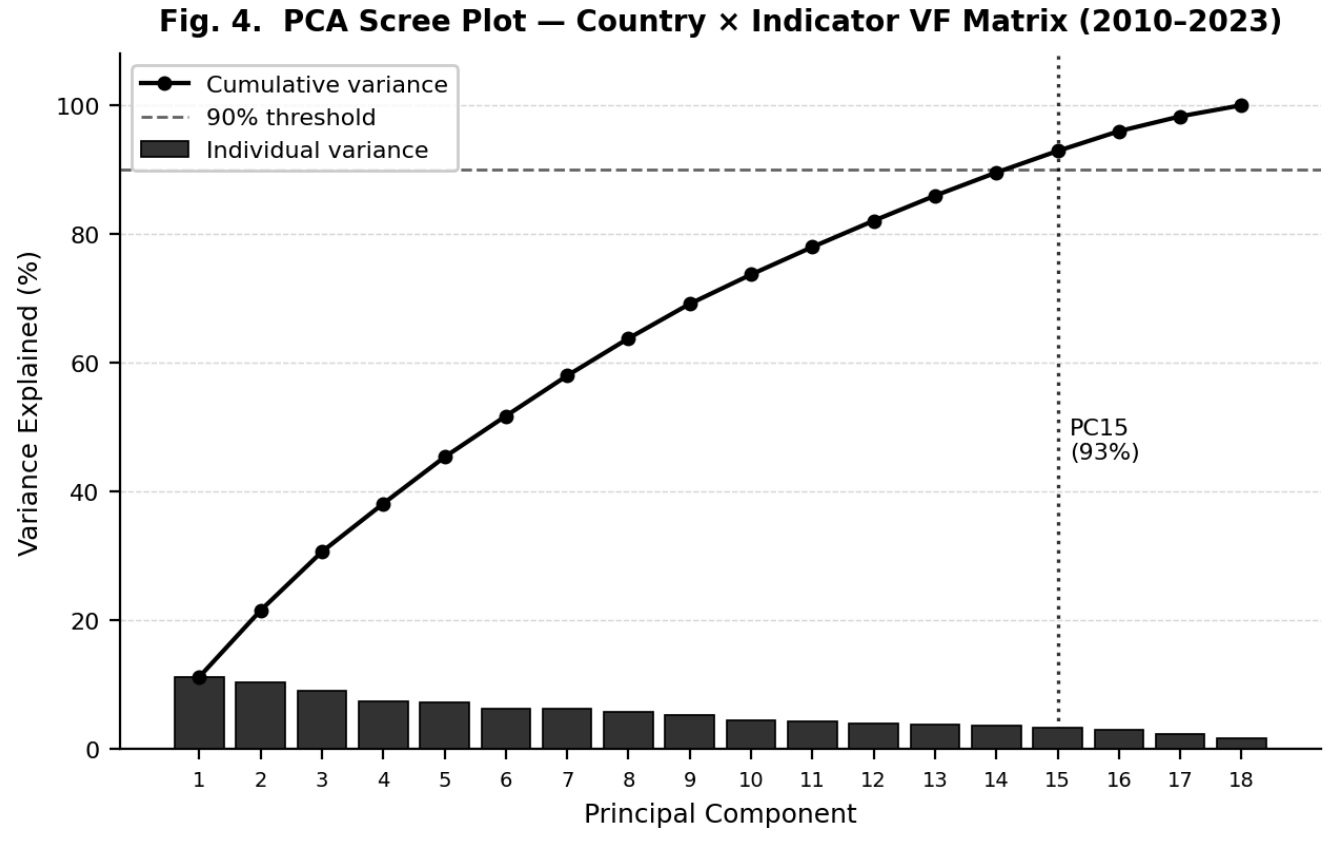

Fig. 4. PCA Scree Plot — Country × Indicator VF Matrix (2010–2023)

PC1 contrasts Regulation and Safety indicators (*AI_Bills*: +0.373; *Train_Compute*: +0.348) against Ethics and Adoption (AI_Nervous: −0.370; WB_GTMI: −0.328). The highest-loading indicators are *AI_Bills* (+0.373), *AI_Nervous* (−0.370), *Train_Compute* (+0.348), and *WB_GTMI* (−0.328), and the component explains 11.3 percent of variance. PC2 shows a mix of Accountability, Security, Transparency, and Ethics indicators, with *AI_Legis* (+0.379), *ICT_Sec_All* (+0.350), *AI_Nervous* (+0.323), and *WIPO_Patents* (+0.315) as the dominant loadings, and explains 10.4 percent of variance. PC3 contrasts Safety against Security, Transparency, and Fairness indicators. *Model_Params* (+0.416), *ICT_Sec_ICT* (−0.404), *AI_Pubs* (−0.358), and *CS_PhD_F* (−0.308) are the principal loadings, explaining 9.2 percent of variance. No single component accounts for more than 11.2 percent of variance, and no thematic dimension loads exclusively on any one component. A PCA biplot of country scores (Figure 5) illustrates this diffuse structure. Individual countries are scattered across the PC1–PC2 plane with no single cluster separating Global North from Global South. These results indicate that geographic accuracy differentials are thematically structured but diffuse. A country's accuracy profile across the 18 indicators cannot be captured by a single geographic bias dimension.

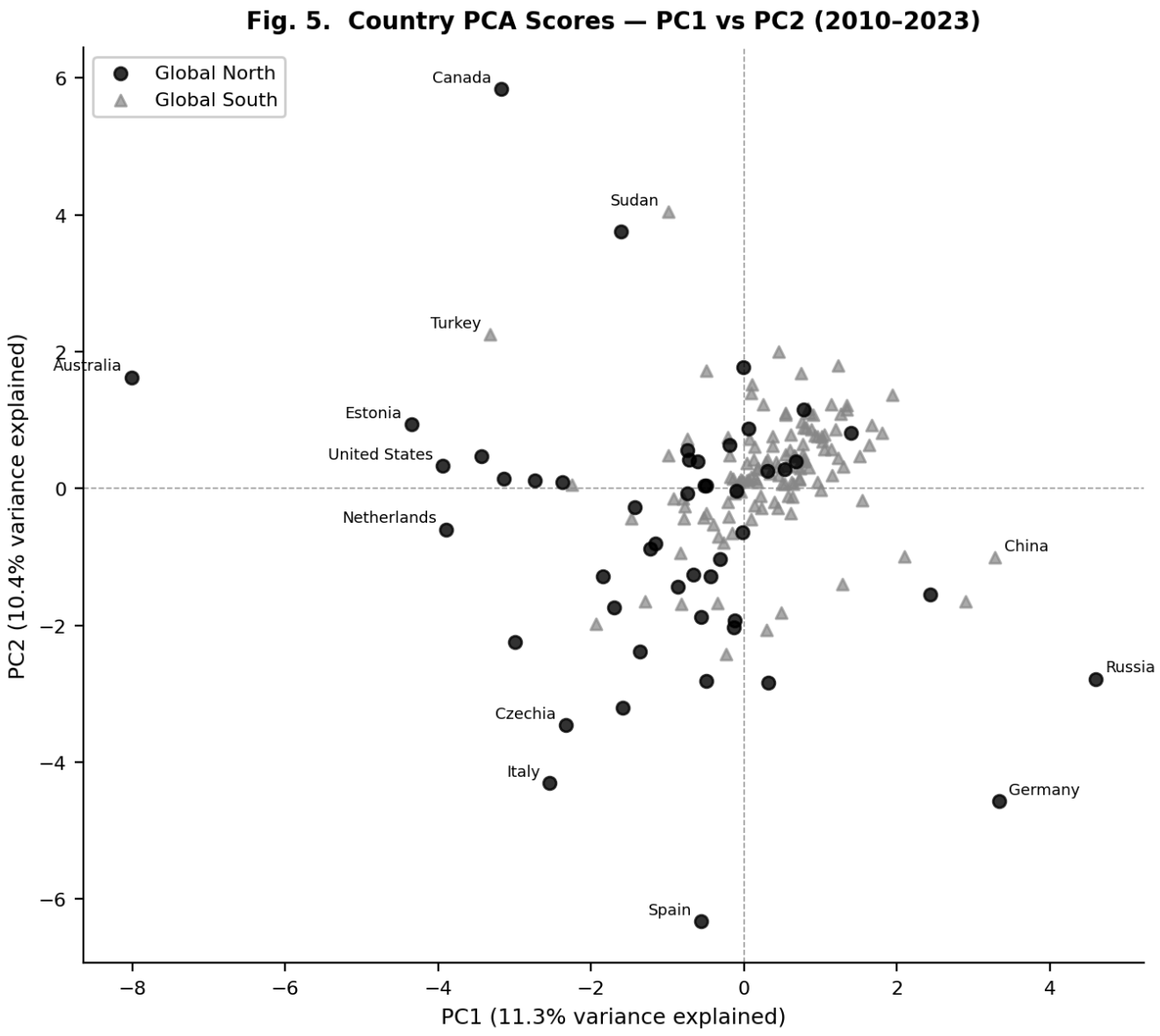

Fig. 5. Country PCA Scores — PC1 vs PC2 (2010–2023)

## G. Temporal Effects (RQ4)

Within the primary analysis window (2010–2019), VF rates were highest in 2010 (30.7 percent) and lowest in 2016 (25.3 percent), meaning fabrication was at its minimum in 2010 and peaked in 2016. The regression confirms a significant increase in fabrication odds between 2010 and 2016 (OR=1.30, $p$=0.009), though fabrication odds declined modestly by 2019 relative to the 2016 peak (OR=1.21 versus OR=1.30 for 2016, $p$=0.068), indicating a non-monotonic rather than linear temporal trend. The year 2013 does not differ significantly from 2010 (OR=1.02, $p$=0.840). In the secondary window (2022–2023), VF rates (24.2 and 27.6 percent, respectively) are consistent with the rates observed for 2016–2019, suggesting that the modest decline in accuracy observed across the primary window does not accelerate further as queries approach the end of the models' training periods. Temporal proximity to training cutoff does not independently explain the geographic accuracy patterns observed, once model identity and thematic dimension are controlled (RQ4: not confirmed).

## VI. DISCUSSION

### *A. The Inverted Geographic Pattern*

The most surprising finding of this study is that fabrication rates are higher for Global North countries than for Global South countries across all four models. This contradicts the dominant prediction in the literature on geographic bias in LLMs, which holds that Global South countries are more susceptible to fabricated responses by virtue of their underrepresentation in training corpora [19], [28].

Two structural features of the dataset and evaluation design most plausibly account for this finding. First, the ±10 percent accuracy threshold used to classify responses as VF is proportional rather than absolute. Countries with small values on a given indicator—a pattern disproportionately observed in Global South countries for metrics such as AI publications, WIPO patent counts, and AI compute capacity—require models to produce proportionally less precise estimates to be classified as accurate. A model responding that a low-income country produced 40 AI publications in a given year, when the verified value is 44, is classified as VF; the same model responding that the United States produced 12,000 publications, when the verified value is 15,000, is classified as HF. The threshold therefore structurally favours accuracy for small-value observations.

Second, the set of country-metric-year observations with verified values in GAID v2—the frame from which this study's 2,990 observations are drawn—is not itself geographically neutral. Many indicators, including AI compute (*Train_Compute, Model_Params*), are populated in GAID v2 almost exclusively for a small number of high-income countries (primarily the United States and China), and it is precisely for these large-value, high-specificity indicators that fabrication rates are highest. The superior accuracy for Global South countries may therefore suggest the composition of the verified-value frame but not an improvement in model knowledge about less-represented countries.

These findings do not imply that geographic bias is absent; they imply that its empirical manifestation, under the evaluation design employed in this study, is more complex than a simple Global North advantage. A more appropriate framing may be that all four models exhibit systematic domain-level fabrication regardless of geography, and that the geographic distribution of this fabrication is shaped as much by the specificity and scale of the verified values as by the geographic prominence of the queried country in training data.

### *B. Thematic Concentration and Diffusion*

The near-total fabrication for Safety indicators is a finding of practical significance. All four models produced numeric estimates for Total Training Compute (measured in FLOP) and Total Model Parameters (measured in billions) that were, in virtually every case, outside the verified range. This outcome is not attributable to geographic bias conventionally (meaning it does not apply equally to queries about compute infrastructure in the United States, China, and the European Union). Rather, the outcome indicates a structural limitation of the factual knowledge encoded in current LLMs regarding the quantitative details of AI hardware deployment. The implication for AI governance analysis is clear. Practitioners who query these models about AI compute capacity should treat any specific numeric response with extreme scepticism.

The high accuracy observed for Regulation indicators—and particularly for the binary National AI Strategy indicator—is also informative. Binary or near-binary indicators (whether a national AI strategy has been published) present a structurally easier task than continuous numeric indicators; a model that knows a country has published a national AI strategy will produce the correct value (1) regardless of the precise date or content of that strategy.

The PCA results indicate that geographic bias, where it exists, does not reduce to a single underlying dimension. A total of 15 components are required to explain 90 percent of variance in the 177-country × 18-indicator accuracy matrix (Figure 4), and the dominant components show mixed thematic contrasts (such as Regulation and Safety indicators in PC1, Accountability, Security, and Transparency in PC2) rather than a simple Global North/South gradient (Figure 5). These empirical outputs support the thematic specificity argument above, where a country's accuracy profile depends more on which indicators are queried and what the verified values are than on a general geographic factor.

### *C. Developer Origin and the DiD Estimate*

The marginally significant DiD estimate (+2.07 percentage points, as shown in Figure 2; $p$=0.076) provides limited support for the proposition that developer national origin moderates geographic accuracy differentials (RQ2). Chinese-origin models (DeepSeek-V3-0324 and Qwen3-235B-A22B) show a somewhat larger accuracy advantage for Global South over Global North queries than Western-origin models (Mistral Large 3 and Llama 4 Maverick), but the effect is small and does not survive correction for multiple comparisons. The developer-origin main effect is negligible (OR=1.003, $p$=0.941), confirming that the two groups do not differ in aggregate accuracy. The evidence from our DiD estimate is insufficient to conclude that developer origin is a substantive moderator of geographic bias, and the estimate may reflect model-specific characteristics that are incidentally correlated with developer geography.

### *D. Query Framing as a Methodological Variable*

Comparative framing (asking how a country performed relative to a regional average) reduced HF odds by 41 percent relative to direct numeric queries (OR=0.59, $p$<0.001), while contextual framing reduced them by 18 percent (OR=0.82, $p$<0.001). This finding suggests that models are less likely to produce confidently incorrect numeric estimates when the query structure invites relative rather than absolute responses. This finding has practical implications for governance analysis, where reframing factual queries in comparative terms may reduce, though not eliminate, the risk of HF.

### *E. Limitations*

Four limitations of this study should be acknowledged beyond those discussed in Section IV-B. First, the proportional ±10 percent classification threshold, while supporting prior work on factual accuracy [10], confounds scale with accuracy in the manner described above. Future studies should consider scale-normalised accuracy thresholds, or separate analyses for indicators with qualitatively different value distributions.

Second, the use of API endpoints for model access (rather than local deployment) introduces a small risk that model behaviour at inference time differs from the published open-weight release. In the case of Llama 4 Maverick and Qwen3-235B-A22B, which were accessed via third-party inference providers, quantised (FP8) deployments were used; quantisation may affect factual precision at the margin. This is particularly relevant for DeepSeek-V3-0324, for which the July 2024 training cutoff is inferred from the model's self-reported system prompt [13] rather than confirmed in the technical report.

Third, the Global North/South classification used in the DiD analysis is a coarse binary variable that fails to detail substantial within-group heterogeneity. A finer-grained analysis using country-level income classifications or regional sub-categories would provide a more differentiated account of geographic accuracy differentials.

Fourth, the absence of HR responses (<0.1 percent) across all four models indicates that contemporary LLMs at frontier scale almost never decline to answer factual queries, regardless of whether they possess the relevant knowledge. This represents a qualitative change from smaller models, which are more likely to refuse queries about less-documented or low-resource countries. The practical consequence is that users of frontier models have no behavioural signal to distinguish HF from confident accuracy.

### *F. Implications for AI Governance Practice*

Three implications follow from these findings for practitioners who use LLMs in AI governance analysis. First, no currently available open-weight frontier model can be relied upon for precise quantitative AI governance data. In our study, even Mistral Large 3, the most accurate model examined, fabricated 61.2 percent of responses. Human verification against primary data sources—such as the Stanford AI Index, WIPO, and the World Bank GovTech dataset—remains essential. Second, the extreme inaccuracy observed for Safety indicators (particularly AI compute metrics) alarms that frontier models should not be used as a primary source for information about national AI compute capacity, model parameter scales, or training resource consumption. Third, the thematic and indicator-level structure of fabrication—which the PCA results show to be multi-dimensional—suggests that accuracy profiles are indicator-specific and cannot be generalised across thematic domains.